# Efficient Computational Design of 2D van der Waals Heterostructures: Band-Alignment, Lattice-Mismatch, and Machine-learning

Kamal Choudhary[1,*1], Kevin F. Garrity[1], Steven T. Hartman[2], Ghanshyam Pilania[2], Francesca Tavazza[1]

1 Materials Science and Engineering Division, National Institute of Standards and Technology,

Gaithersburg, Maryland 20899, USA

2 Materials Science and Technology Division, Los Alamos National Lab, Los Alamos, New Mexico 87545, United States.

## Abstract

We develop a computational database, web-apps, and machine-learning (ML) models to accelerate the design and discovery of two-dimensional (2D)-heterostructures. Using density functional theory (DFT) based lattice-parameters and electronic band-energies for 674 non-metallic exfoliable 2D-materials, we generate 226,779 possible heterostructures. We classify these heterostructures into type-I, II and III systems according to Anderson's rule, which is based on the relative band-alignments of the non-interacting monolayers. We find that type-II is the most common and the type-III the least common heterostructure type. We subsequently analyze the chemical trends for each heterostructure type in terms of the periodic table of constituent elements. The band alignment data can be also used for identifying photocatalysts and high-work function 2D-metals for contacts. We validate our results by comparing them to experimental data as well as hybrid-functional predictions. Additionally, we carry out DFT calculations of a few selected systems ($MoS_2/WSe_2$, $MoS_2$/h-BN, $MoSe_2/CrI_3$), to compare the band-alignment description with the predictions from Anderson's rule. We develop web-apps to enable users to virtually create combinations of 2D materials and predict their properties. Additionally, we use ML tools to predict band-alignment information for 2D materials. The web-apps, tools and associated data will be distributed through JARVIS-Heterostructure website (website: https://jarvis.nist.gov/jarvish/). Our analysis, results and the developed web-apps can be applied to the screening and design applications, such as finding novel photocatalysts, photodetectors, and high-work function (WF) 2D-metal contacts.

[1] Corresponding author: kamal.choudhary@nist.gov

## Introduction

Heterostructures are interfaces between dissimilar materials and are of practical importance in a wide variety of optoelectronic devices[1]. Two-dimensional (2D) materials exhibiting van der Waals (vdW) bonding strictly in one of the three crystallographic directions and their heterostructures provide a huge variety of functionality[2, 3] and property-control[4, 5] because their surfaces are generally free from dangling-bonds, extended defects and trap states[6] (which is in stark contrast with semi-coherent and incoherent heterointerfaces found in three-dimensional crystalline solids). Additionally, they allow integrating various nanoscale materials to create diverse vdW heterostructures because they do not require strict constraints of crystal lattice matching[6] and have a wide variety of bandgaps. Despite their widespread applications, a systematic high-throughput investigation of a wide range of 2D chemistries is still lacking, mainly because of the overwhelmingly large number of underlying combinatorial possibilities. The number of possible 2D materials could be in the thousands as shown by recent computational screenings[7-11] , so there can be millions of potential pair combinations of distinct heterointerfaces. Moreover, since experimental routes for investigating these heterointerfaces usually are labor and resource-intensive, the use of computational methods, such as density functional theory (DFT), to perform screening is a very well justified first step in an elaborate materials design process.

Metallic 2D materials with high work-functions[12] are useful for battery-applications[13] as well as minimizing Schottky barrier[14], while 2D semiconducting or insulating materials are good candidates for making different types of hetersostructures[15] such as type-I (symmetric), type-II (staggered), and type-III (broken). We determine the types using Anderson's rule, which considers the conduction and valence band information of constituting monolayers only. The information about Conduction Band Minima (CBM) and Valence Band Maxima (VBM) can easily be obtained from DFT. A type-I band alignment is most widely utilized in optical devices, such as light-emitting diodes (LEDs)[16] and lasers. A type-II band alignment is useful for unipolar electronic device applications, including photodetectors,[17-19] as it can allow for more efficient charge separation and rapid transport for valence and conduction band carriers. The type-III heterostructures are used for tunneling field-effect transistors(TFETs)[20]. Magnetic/non-magnetic heterostructures are also very interesting for proximity-effects during phenomena such as band-tuning[21] and valleytronics[22]. Other important applications of 2D heterostructures include self-cleansing[523], plasmonic devices[24], electroluminescence[25], complementary metal–oxide–semiconductor (CMOS)[26], and DNA biosensors[6, 27].

There has been substantial research in the vdW heterostructure design, especially for graphene, BN, black-Phosphorous, and chalcogenides. Some of the highly investigated heterostructures using experimental methods are: $WSe_2/MoS_2$[19], Graphene/$MoS_2$[27], Graphene/$WS_2$[17], $MoS_2$/h-BN/graphene[28] and $WSe_2/CrI_3$[22]. Some of the notable computational works for 2D heterostructure-design include the ones by Ozcelik, et al.[15], Chiu et al.[29] and Wilson et al.[30], and Andersen et al.[31]. However, the next generation of 2D materials has gone beyond the known classes and includes many more chemical and structural classes than those investigated in the computational works mentioned above. For a systematic study of heterostructures, a robust algorithm for constructing heterostructures using lattice parameters information and a database of 2D materials' structure and electronic information is essential. Fortunately, recently several computational tools have been developed to identify best-matched configurations including works by Mathew et al.[32], Dwarkanath et al.[33] based on the algorithm proposed by Zur and McGill[34]. Also, there have been several 2D materials databases that host CBM, VBM, vacuum potentials, and Fermi-energies for 2D materials, such as the JARVIS-DFT[35-39] database, C2DB[40], and MARVEL[8]. The easy availability of all the necessary tools for interface design, computational data for 2D materials, and experimental data for validation allows us to investigate a vast majority of all possible combinations of heterostructure chemistries.

Experimentally, 2D-heterostructures are created either using manual stacking or via direct synthesis using, for instance, chemical vapor deposition (CVD) processes. Unlike direct CVD growth, whereby vdW heterostructures typically adopt a certain fixed orientation, the manual stacking approach offers a new degree of freedom to tailor the relative twist angle (within an error of $1°$)[41] between different layers in nearly arbitrary configurations. Moreover, the heterostructures can be created as vertical stacks or as laterally stitched junctions. We consider only vertical junctions in this work. Although 2D heterostructures may not require a strict lattice matching criterion, the presence of sufficient lattice mismatch along with weak vdW bonding between the 2D layers can lead to incoherent lattice matching, and the generation of Moiré patterns[42].

In this work, we aim to provide a sufficiently accurate and highly efficient model to preselect 2D heterostructures for specific applications among a vast pool of candidates. We develop a database and an open-access web interface to facilitate the generation of 2D heterostructures and their classifications. We use the method of Zur et al.[33, 34] to generate interfaces between these 2D materials, and to calculate lattice constants and angle mismatches. The set of 2D materials is obtained from the JARVIS-DFT 2D database, from which we extracted the VBMs, CBMSs, Fermi-energies, equilibrium structures, and wavefunction data. JARVIS-DFT is a database containing about 38000 bulk and 900 low-dimensional materials with their DFT-computed structural, exfoliability[35], elastic[36], optoelectronic[37] solar-cell efficiency[39], and topological[39] properties. It is part of National Institute of Standards and Technology (NIST) effort within the scope of Materials Genome Initiative (MGI) related activities. Next, we develop an easy-to-use web-app for predicting the heterostructure type and lattice mismatches between all the possible 2D materials in our database. As mentioned above, Anderson's rule-based designation of heterostructure type requires only the monolayer CBMs and VBMs of the monolayers forming the heterostructure. Since DFT calculations for a vast number of new 2D materials could still be computationally expensive, we leverage machine-learning (ML) techniques to further accelerate the predictions of CBMs and VBMs of 2D materials given the crystal structure information. We train ML models for CBM, VBM and work-functions of 2D materials using the Atomistic Line Graph Neural Network (ALIGNN) method[43]. We believe that the easily available heterostructure database along with other calculated properties would serve as a useful tool for the community towards novel and application-specific 2D-heterostructure design.

## Results and discussion

We start by employing available JARVIS-DFT data for 2D materials to predict the nature of 2D-heterostructure band alignments. A flow-chart associated with this process is shown in Fig. 1.

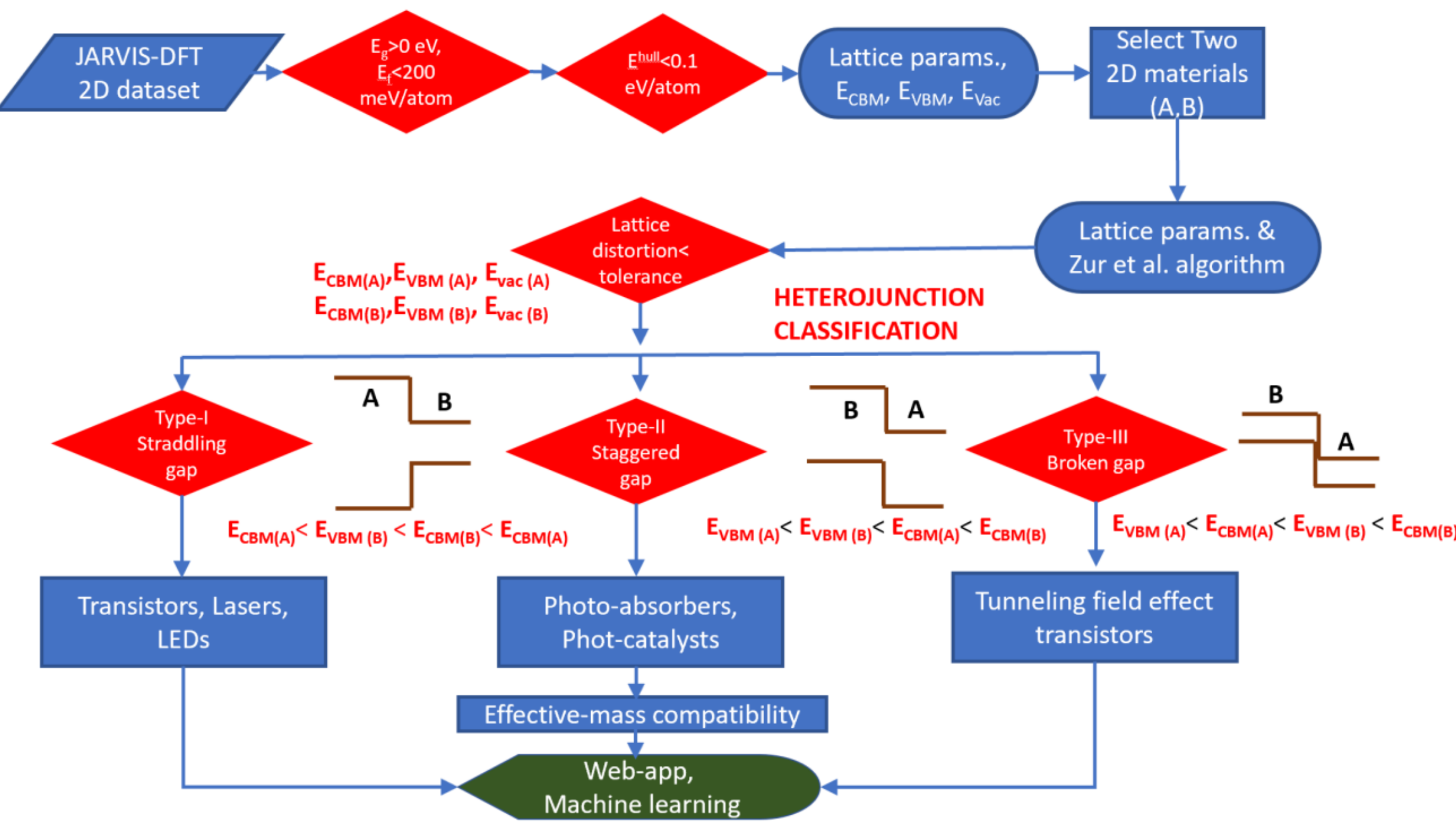

*Fig. 1 Flowchart for the computational design of 2D-heterostructures.*

The JARVIS-DFT hosts a) DFT-optimized structures and electronic structure information such as b) CBM, c) VBM, and d) vacuum level information (VAC). For a heterojunction, we align the 2D materials VBMs and CBMs with respect to corresponding vacuum levels. First, we select non-metallic ($E_g$ > 0.0) 2D materials with exfoliation energy less than 200 meV/atom and their bulk counterparts with energy above convex hull less than 0.1 eV/atom, leading to 674 2D materials.

The lattice parameter data is then used to find the best match within the heterointerface twist angle tolerance of $1^{o}$ and mismatch of 0.05 % using the algorithm developed by Zur et al.[33, 34], which finds the heterostructure configuration with the minimal mismatch. Note that we avoid highly mismatched heterostructures because they may lead to the generation of dislocation, grain-boundaries and internal-stress, interface buckling and other effects that are beyond the scope of this work. These heterostructures can be now classified as type-I, type-II, or type-III structures based on the VBM and CBM data mentioned above based on Anderson's rule[49]. A heterojunction is type-I if VBM(A)<VBM(B)<CBM(B)<CBM(A), type-II if VBM(A)<VBM(B)<CBM(A)<CBM(B), and type-III if VBM(A)<CBM(A)<VBM(B)<CBM(B), assuming that the lowest VBM material is always labeled as A.

It is possible that explicit DFT calculations of the heterostructure are needed to generate an accurate density of states(DOS). Therefore, to assess the accuracy of our predictive approach, we compare the DOS of individual 2D materials, their heterostructure and predicted band-alignment diagrams for a selected set of example materials in Fig. 2. We find that both the $MoS_2$-$WSe_2$ (JVASP-664-JVASP-652) (Fig. 2a-d), $MoS_2$-BN (JVAS-664-JVAS-688)- (Fig. 2e-h) and $MoSe_2$-$CrI_3$ (JVASP-649-JVASP-76195) (Fig. 2i-l) interact feebly, hence the band-alignments predicted using Anderson's rule are similar to those obtained with the DFT computations. For instance, the DOS of Mo and S are almost unchanged between pure $MoS_2$ and $MoS_2$ as part of $MoS_2$-$WSe_2$ heterostructure (Mo and S DOS in Figs. 2a and 2c are similar). A similar finding applies to both the W and Se DOS. In the resultant heterostructure, the net gap of the system is smaller than both bandgaps of individual materials, as shown in the band-diagram (Fig. 2d), due to the offset of the two materials' gaps. Similar behavior is also observed for the $MoS_2$/h-BN and $MoSe_2$-$CrI_3$ heterostructure.

For magnetic/magnetic, magnetic/non-magnetic 2D material interfaces, we predict the band-alignments based on net CBMs and VBMs of both spin channels. Note that the electronic structure of heterostructures with internal stress due to mismatch could be different from that of the two constituent materials, and it would be difficult to predict using the Anderson rule. Ozcelik, et al.[15] reported such behavior for $MoSe_2$-$WSe_2$ heterostructure. These results also assume that the two 2D-materials do not interact much. However, charge transfer could be possible especially for low-bandgap materials which would change the band-alignment. Also, note that we discuss intrinsic, defect-free and 0 K results in the current work, while at ambient conditions, equilibrium thermodynamic concentration of defects in systems, and doping levels could also significantly change the heterostructure behavior.

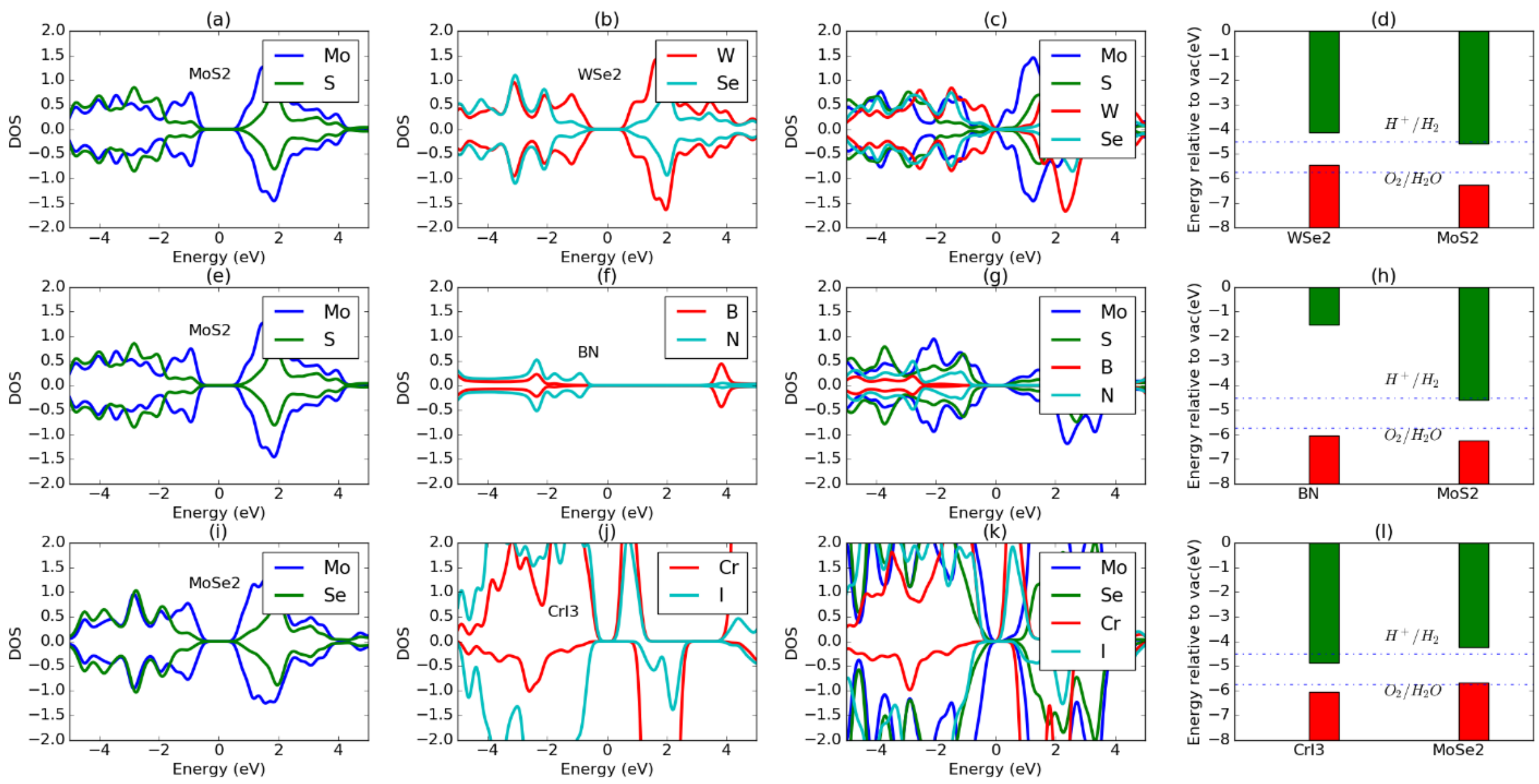

*Fig. 2 (a) DFT calculated DOS for a monolayer of $MoS_2$ in vacuum, normalized to the Fermi energy of the monolayer. (b) The same quantity for $WSe_2$. (c) The combined DOS of the bilayer $MoS_2$/$WSe_2$ heterostructure in vacuum, normalized to the Fermi level of the heterostructure. (d) $MoS_2$/$WSe_2$ band alignment diagram, with energy levels for each material calculated in vacuum* *and aligned relative to the vacuum energy level. (e-h) The same quantities for $MoS_2$/h-BN, (i-l) The same quantities for $MoSe_2/CrI_3$.*

In Fig. 3, we predict the types of all the 226,779 heterostructures that were obtained by all possible combinations within the specified lattice mismatch tolerance. Currently, the number of known 2D-heterostructures is in the range of hundreds or thousands, this work suggests thousands to millions of possible heterostructures based on simple combination rules. Each dot in Fig. 3a represents a heterostructure. Fig. 3a shows that it is possible to obtain a variety of heterostructure types for most of the 2D materials. Some materials, however, show a clear preference for a specific heterostructure type. This type of preference is revealed by the appearance of stripes in the Fig. 3a. The stripes are visible mainly for green and red, i.e., type-I and type-III heterostructures. As shown in the Fig. 3b, we find that the majority of the heterostructures are of the type-II (≈37.8 %), the second most abundant type is the type-I (33.4 %), and rarest are of type-III (28.8 %).

We investigate the overall probability that a specific element in the periodic table would form each heterostructure type. Our results are shown in Fig. 4. In order to understand the elemental contributions, we weight an element in both the 2D-systems forming the heterostructure as one or zero separately for the three heterostructures classes (i.e., type-I, II or III heterostructures). After performing such weighting for all the materials in our database, we calculate the probability that an element is part of a particular type of heterostructure. Suppose there are x number of Se-containing materials and y of them form type-I heterostructure then the percentage probability (p) for Se in type-I is calculated using the formula: $p = \frac{y}{x} \times 100$ %. Interestingly, we find that type-I and type-III heterostructures tend to form with specific elements, while type-II heterostructures are possible with almost all the elements in the periodic table. B, Y, Ru are clearly highly-probable for type-I, Rh, Ir, Pt and the group III onwards elements for type-II, and alkali elements and Mn, Fe transition metal elements tend to form type-III heterostructures. Although the Periodic Table trends provide an overview of the elemental probability, the trends are not obvious and hence there is a need for actual DFT predictions or experiments to confirm the type of the heterostructure. Below we discuss these individual types in further detail and focus on some specific examples where the identified materials can be potentially useful from an application point of view.

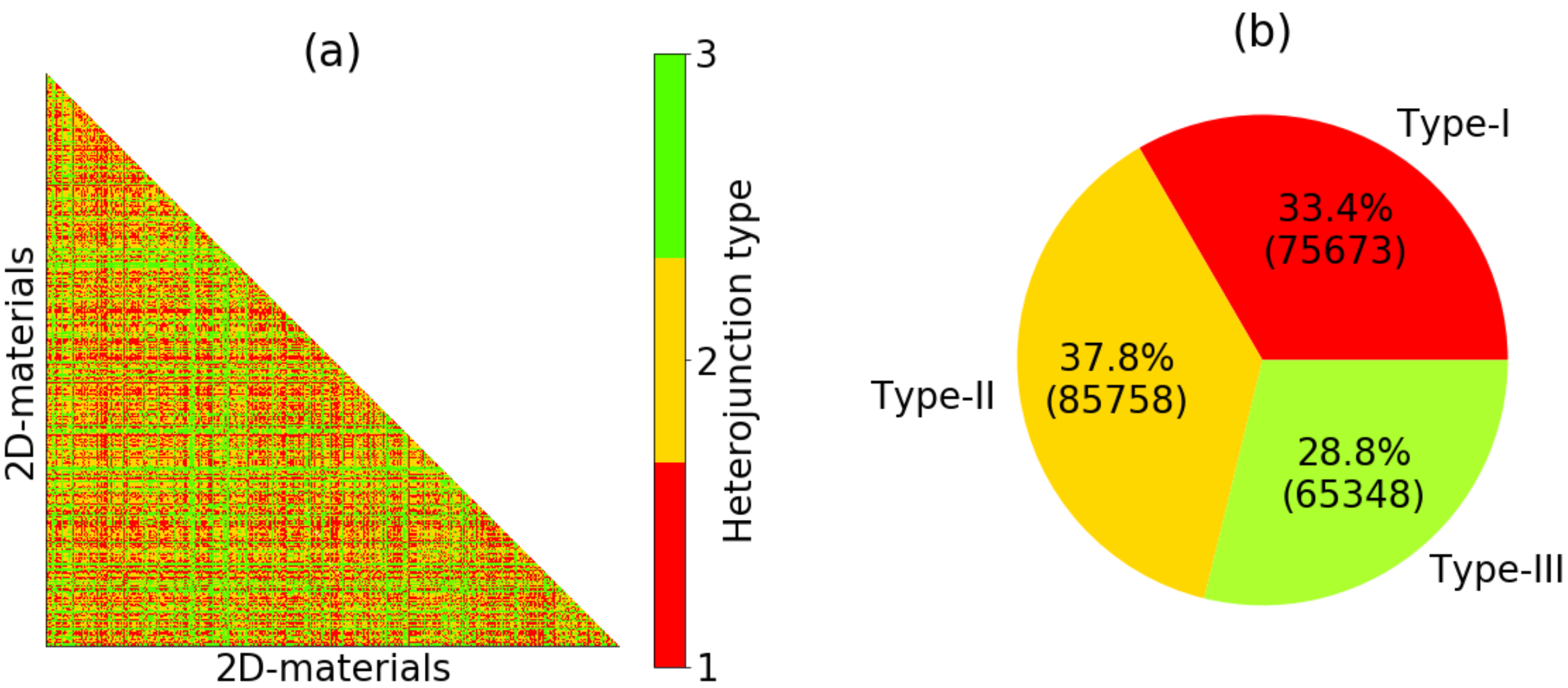

*Fig. 3 Distribution of type-I, type-II and type-III heterostructures. a) Combination of all possible heterostructure type for a 2D material. Each point in the plot represents a heterostructure. Each point on x and y axes represents a 2D material. Only the lower diagonal is plotted because it is a symmetric matrix. b) A pie chart of the type-distribution for the three heterostructures.*

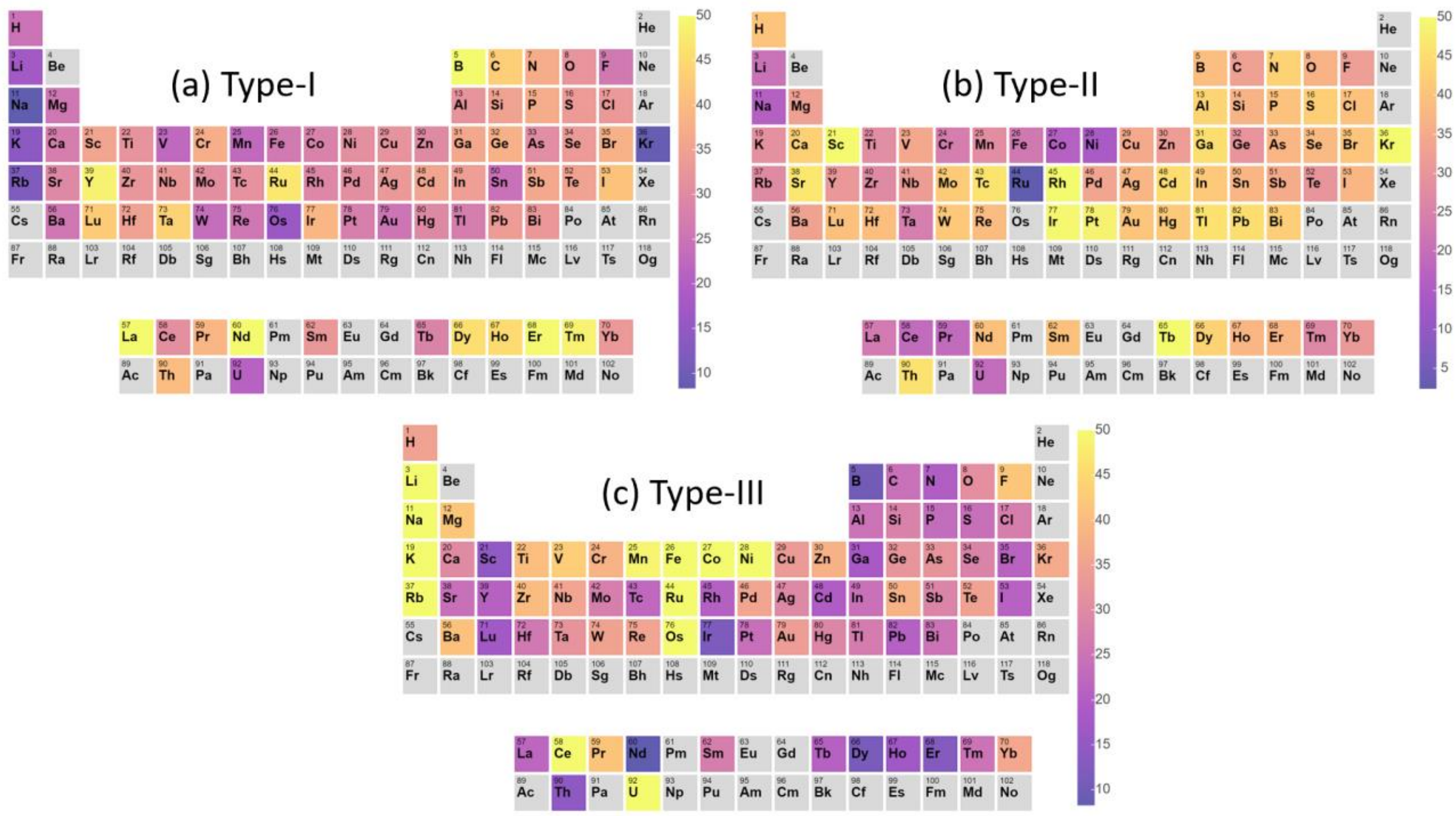

*Fig. 4 Periodic Table trends of percentage probability that an element is a part of type-I, type-II and type-III heterostructures.*

**Type-I heterostructure and optoelectronics applications**

Type-I heterostructures are used in transistors, light-emitting diodes (LED) and laser applications[16]. In this case, both the CBM and VBM of the heterointerface are located in the material with narrower bandgap. The quantum confinement of electrons and holes in the same region facilitates their radiative recombination, which is desirable in light-emitting applications. An example of such experimentally-confirmed type-I materials is $MoS_2$-$ReS_2$[50], where $ReS_2$ has a narrower gap than $MoS_2$. Using the CBM and VBM datasets, we also find the heterostructure to be type-I, in agreement with the experimental results (shown in the band-alignment diagram in Fig. 5a). Our methodology predicts 21 other 2D-materials that could form type-I heterojunctions with 2H-$MoS_2$, such as $ZrSe_3$ (JVASP-6019), $PtSe_2$(JVASP-744), $WS_2$(JVASP-652), $Bi_2Se_3$(JVASP-6736), $InAg(PS_3)_2$ (JVASP-6358), SbSI (JVASP-6724), and $As_2S_3$ (JVASP-6451).

Therefore, more generally, the computational-tools developed in this work can easily predict all possible potential heterostructures for a given 2D-materials and a specific heterojunction type. Many of these predicted heterostructures are not reported experimentally to the best of our knowledge. Hence, the screening work performed within this study could help guide future experiments towards desired heterostructure types.

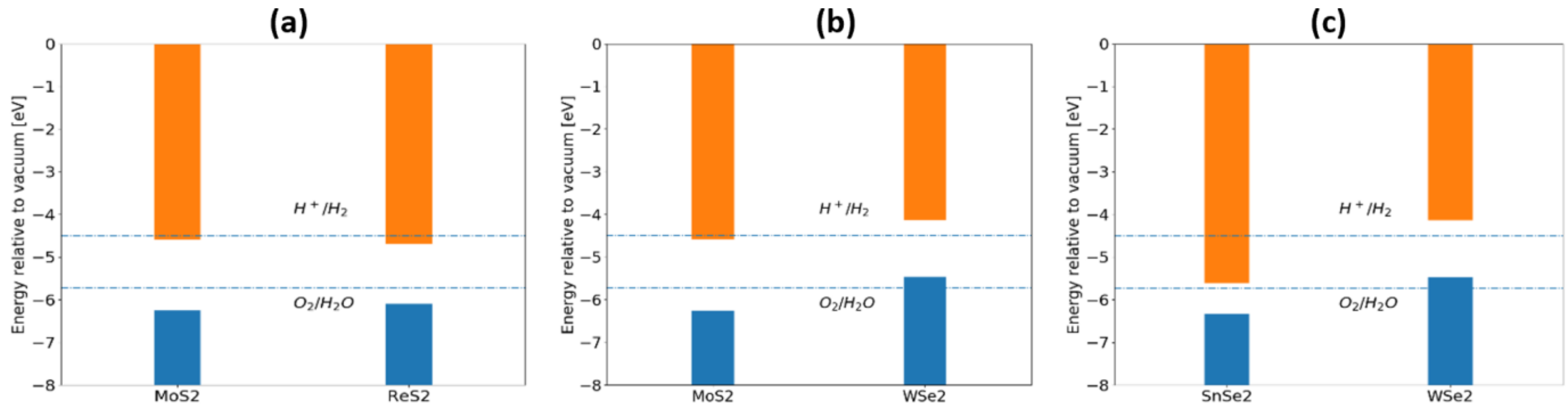

*Fig. 5 Examples of type-I, type-II and type-III heterostructure band-alignments relative to vacuum energy. a) 2H-$MoS_2$/$ReS_2$, b) $MoS_2$/$WSe_2$, and c) $SnSe_2$/ $WSe_2$ interface. The dotted lines show the $H^+/H_2$ and $O_2/H_2O$ levels that must be bridged by the interface CBM/VBM for water-splitting photocatalysis.*

## Type-II heterostructure and photodetection/absorption applications

In type-II materials, the CBM and VBM of the resultant structure are located in two different materials forming the interface. The excitation of the wide-gap material is followed by the transfer of electrons, but not the holes. The opposite charge transfer occurs when the narrow-gap material is excited. The separation of the electrons and holes to different layers can increase their lifetime and is desirable for photovoltaics and photodetection applications. A well-known example of type-II 2D heterostructure is $MoS_2$/$WSe_2$ interface[51]. Using our methodology, we predicted this system to be type-II as well, which once again validates our approach (Fig. 5b). The CBM resides in $MoS_2$ while the VBM in $WSe_2$. In this case, upon irradiation, photons are absorbed, and excitons are generated in the single-layers of $WSe_2$ and $MoS_2$. The photo-generated free electrons in the CBM of $WSe_2$ can be transferred to the CBM of $MoS_2$ owing to the large band offset between $WSe_2$ and $MoS_2$. Therefore, compared to the single-layer materials, the heterojunction leads to a better utilization of light and provides more electrons for the reaction, therefore increasing PEC current up to ~6 times compared to single 2D materials[51]. As an example, we predict 43 2D materials that could form type-II heterojunction with 2H-$MoS_2$ including GaTe (JVASP-6838), $MoSe_2$ (JVASP-649), $CrBr_3$ (JVASP-6088), $Sb_2S_3$ (JVASP-6523).

### Type-III heterostructure and TFETs

Lastly, in type-III heterojunctions, the band-edges do not overlap at all, making the overall heterojunction metallic. The situation for carrier transfer is like type-II, just more pronounced. Type-III heterojunctions are used in TFETs. TFETs could be used to mitigate the issue of conventional metal–oxide–semiconductor field-effect transistors (MOSFETs) due to scaling down the supply voltage with the increasing power density and the thermionic emission limitation of carrier injection[52, 53]. Unlike type-I and II, there are very few type-III heterojunctions known[52]. An experimentally known example of type-III heterojunction is $SnSe_2$/$WSe_2$[52](JVASP-762, JVASP-652) which we also find as type-III (Fig. 5c). As an example, screening for 2D materials that can form type-III with 2H-$MoS_2$, we find $MnO_2$ (JVASP-6922). Similarly, searches for al l the possible combinations could be performed to guide computations or experimental work.

### Water-splitting applications

2D-materials can be used for water-splitting to generate $H_2$ and $O_2$ gases when irradiated with sunlight[54, 55]. For a material to be able to split water, its CBM must lie at a more negative potential than the reduction potential ($H^+$/$H_2$: 0.0 eV vs. Normal hydrogen electrode (NHE) at pH 0) and the valence band maximum must exhibit more positive potential than the oxidation potential ($O_2/H_2O$: +1.23 eV vs. NHE at pH 0). Out of all the 2D materials in our database, we could find 180 2D monolayer materials that satisfy this condition as shown in Fig. 6. Some of the materials are already known water splitters, including BN[56], black-P[54], ZnPS3[57]), but many other materials, including $As_2O_3$, InSe, HfBrN, and $ZnSiO_3$ are not known experimentally as photocatalysts for water-splitting. As the band-edges could be tuned using different types of 2D materials, the heterostructures would allow a greater number of materials with water-splitting capabilities. Also, a heterostructure, especially of type-II as discussed above, can prevent electron-hole pairs (EHP) from recombining by separating the electrons and holes, which would increase the photogeneration rate.

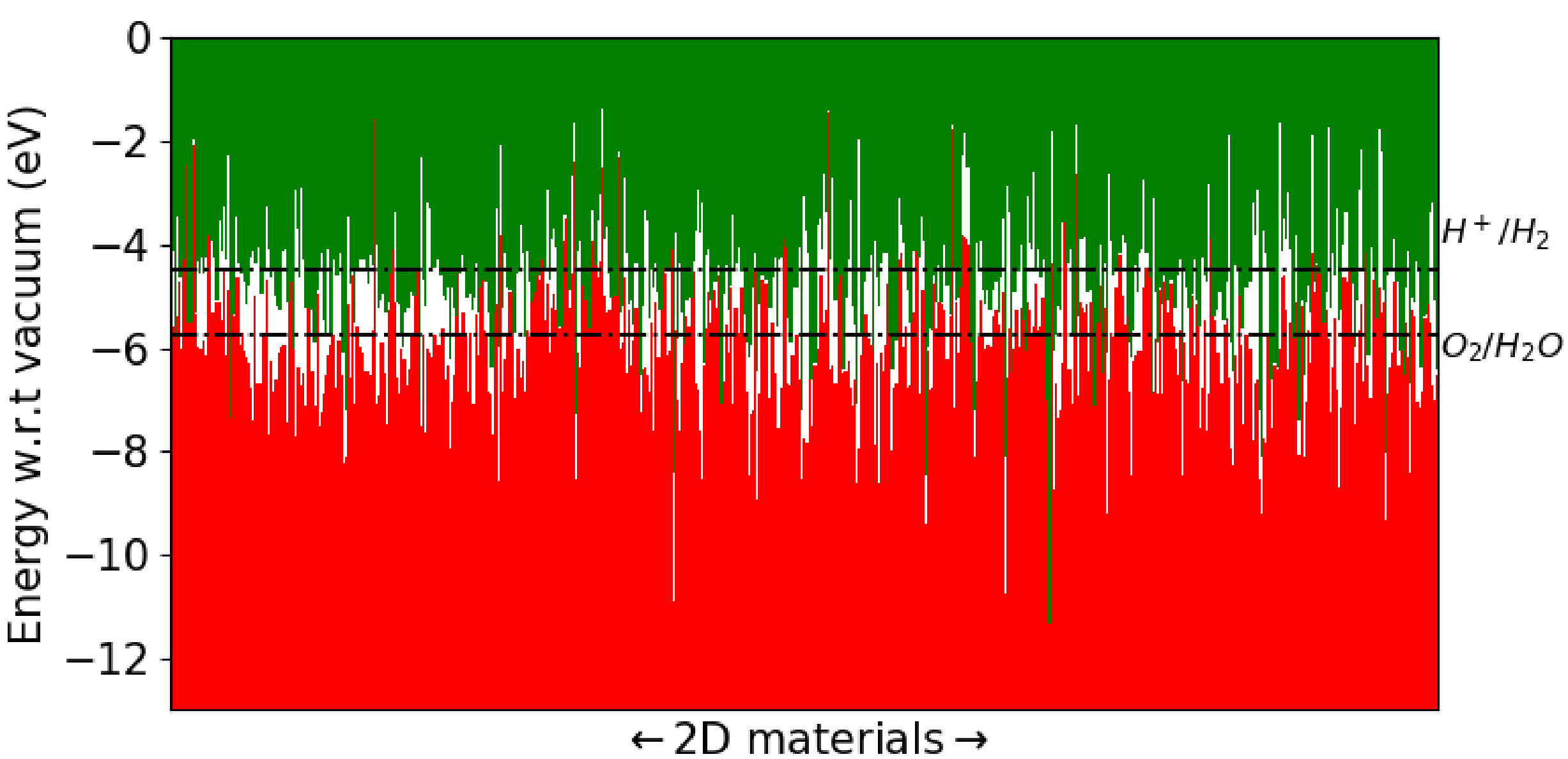

*Fig. 6 Band-alignment of the non-metallic 2D materials with respect to $H^+/H_2$ and $O_2/H_2O$ levels. VBMs are drawn in red while CBMs are green.*

## High work-function 2D-materials

Heterostructures, which utilize 2D materials as metallic contacts instead of conventional metallic contacts such as Al, Pt and Cu are of great importance in designing electronic devices. But very few such 2D materials with high WF have been reported[12-14]. These materials have been shown to act as an excellent contact for 2D heterostructures and could be advantageous compared to conventional contacts such as Pt because of easier lattice mismatch. In Fig. 7, we show the WF distributions for all the 2D metals. We observe that most of the WFs are centered around the 4 eV to 5 eV window, but high WFs are also possible. Some of the high work-function 2D-metals are: $WO_2$ with 9.6 eV (JVASP-783), $MoO_2$ with 9.3 eV (JVASP-78036), $TiO_2$ with 8.77 eV (JVASP-765), $ZrN_2$ with 7.07 eV (JVASP-75043), $PtCl_2$ with 6.96 eV (JVASP-28136) etc. As expected, most of the high WF 2D materials are oxides.

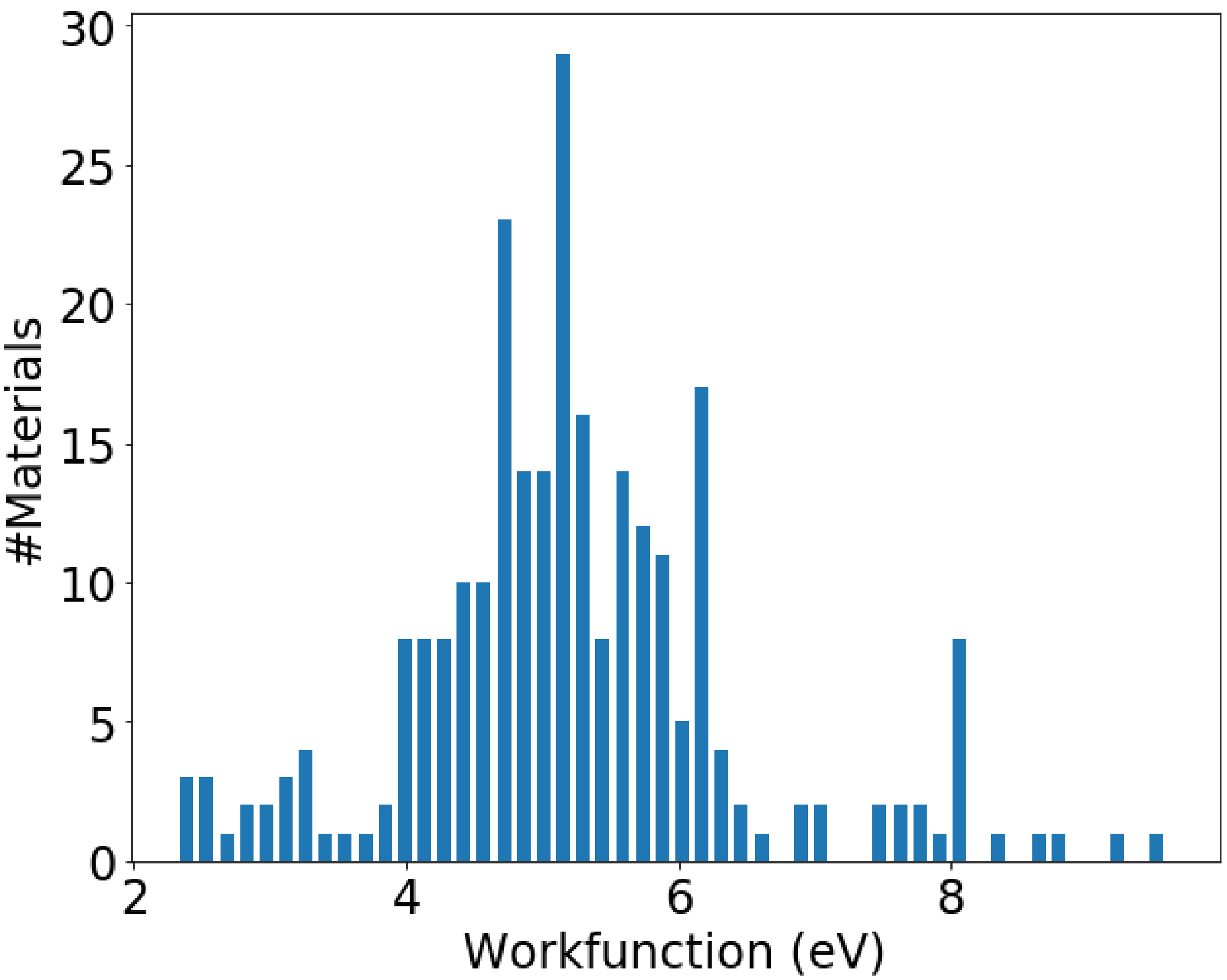

*Fig. 7 WF distribution of 2D metals from the JARVIS database.*

**Web-app development**

We now describe the details of the developed website application (web-app), which is specifically designed to facilitate easy predictions and further analysis of 2D-heterostructure. A snapshot of the *JARVIS-Heterostructure web-app* is shown in Fig. 8. In this app, the electronic bands data and lattice-matching algorithm are used to predict the possible 2D heterostructure and its type. Currently, there are two sub-apps: a) a user can select options of two 2D materials (for which computed data is already available for 2D materials) to predict the lattice-mismatch and band-alignments of the heterostructures, b) computationally design a heterostructure for arbitrary 2D materials. The possible options are all the semiconducting 2D materials in the database. In addition to the mismatch and band-alignment information, the app shows the band-diagram to help visually analyze the results. In case some material is not in the database, the user can put two 2D materials crystallographic information in the second app, which then makes the 2D heterostructure under the pre-defined mismatch criteria. The user can then run the calculations on the three structures to predict the heterostructure properties. The code to generate band-alignments and the heterostructure is freely available at the jarvis-tools github page (https://github.com/usnistgov/jarvis). The web-app will be made available at: https://jarvis.nist.gov/jarvish/ .

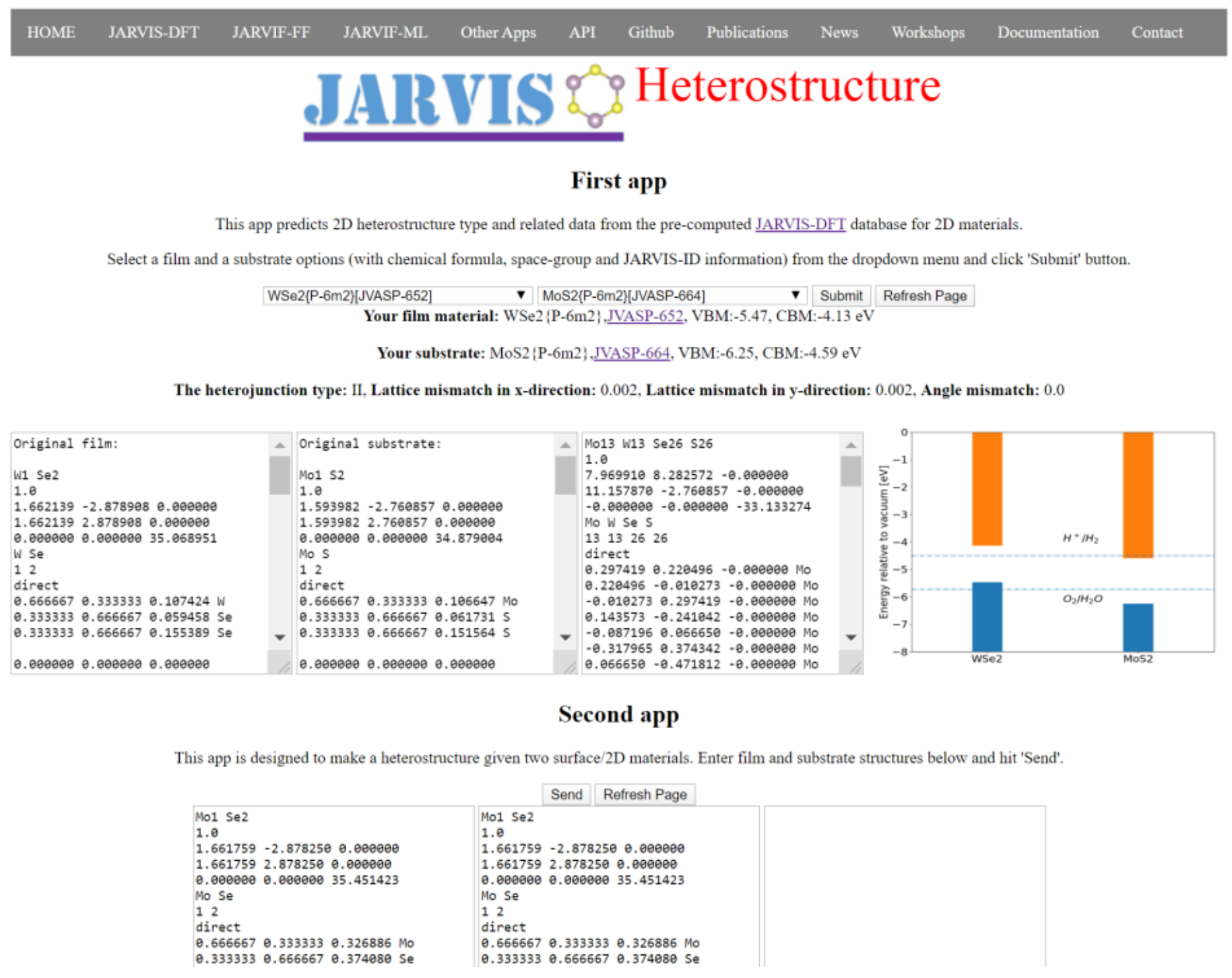

*Fig. 8 Snapshot of the JARVIS-Heterostructure app showing two sub-apps.*

## Machine-learning applications

To treat possible 2D materials that are not yet in our database, we develop machine-learning (ML) models for the band diagrams of 2D materials, using the ALIGNN method[43]. The data are taken from the JARVIS 2D-dft dataset, for a total of 1,062 2D materials. We use 80% of the data for training, 10% for hyperparameter optimization, and 10% for test scoring; hyperparameters and other training details are provided in the Methods section. We consider CBMs, VBMs, WFss separately from the JARVIS-DFT calculation results as the target data for ML. In the Fig. 9a-c we visualize the target data distribution. This helps us understand the boundaries in which the ML algorithm would work well, as ML algorithms are good for interpolation purposes only and generally fail at extrapolation. The results of this test are shown in Table. 1. As it is possible that a 80/10/10 split could be biased; therefore, we perform five-fold cross-validation on each model with complete data. We report the corresponding mean absolute error and standard deviations as shown in the Table. 1 as well. In Fig. 9d-f we show the DFT vs ML predictions after training the ALIGNN model on 80 % training data and testing on 10 % test set data. The results for the trained models are given in the Table. 1 and clearly suggest that for most of the models we achieve reasonable-accuracy compared to the average-predicting baseline model and the models could be used on new materials which are not currently available in our database for a first line of screening. On the second app discussed above, we plan to integrate the ML models so that users can input crystal structure and find the best-matched interface as well as get ML predictions for the heterostructure design.

*Table. I Machine-learning model performance using Mean Absolute Error (MAE) as a metric. Mean and standard deviation values are shown for 5-fold cross-validation (CV) results also.*

| **Model** | **MAE (Test-set)** | **Baseline MAE** | **5-fold CV MAE** |
| --- | --- | --- | --- |
| **VBM (eV)** | 0.47 | 1.47 | 0.41±0.09 |
| **CBM (eV)** | 0.47 | 1.07 | 0.40±0.10 |

| **Work-function (eV)** | 0.23 | 0.95 | 0.41±0.10 |
|---|---|---|---|

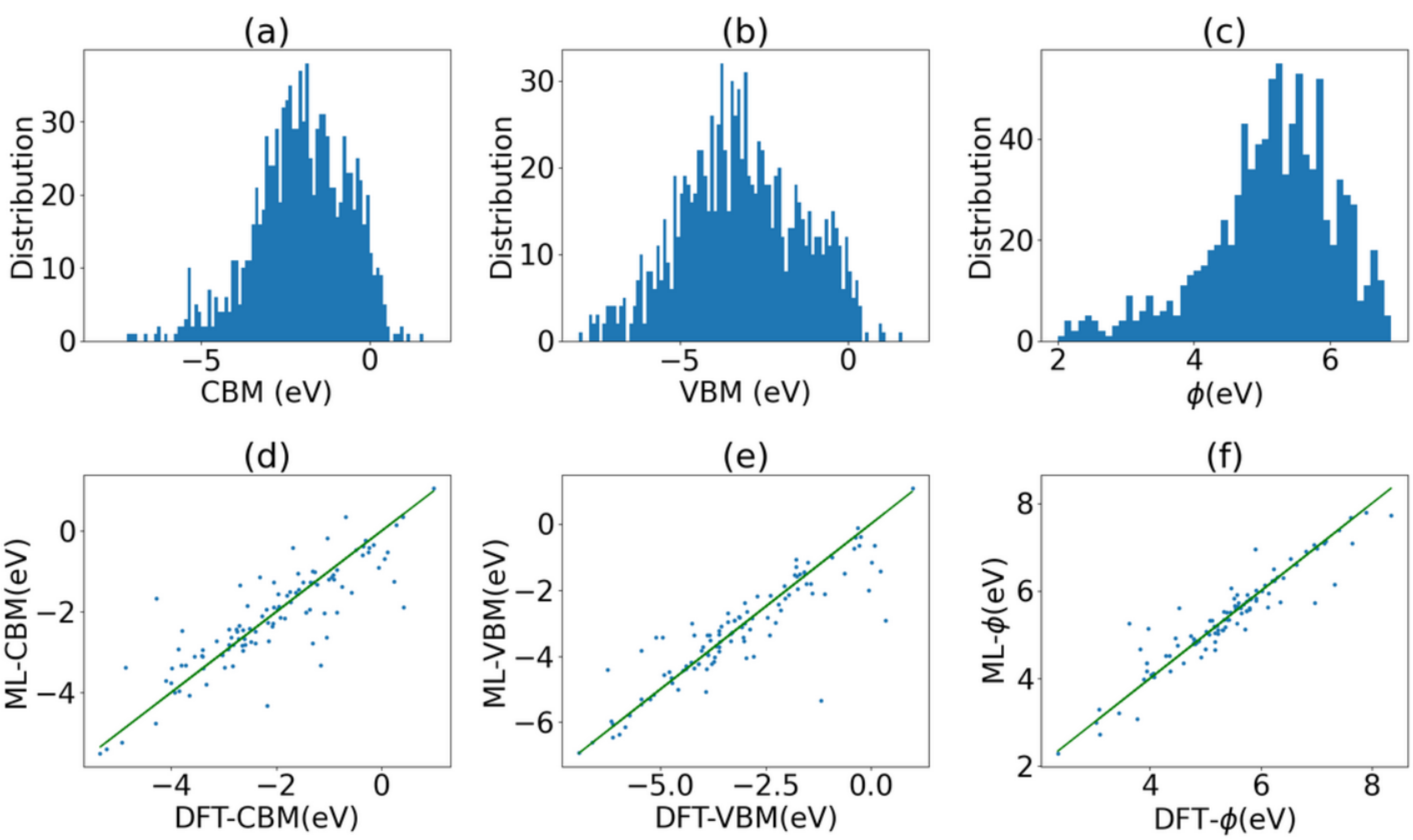

*Fig. 9 (a-c) Histogram of the JARVIS 2d materials data for CBM and VBM, relative to the vacuum level, as well as WF. (d-f) Parity plots comparing the DFT-calculated values for these properties on the 10% test set, to the values predicted with ALIGNN ML method. The green line at y=x is a guide for the eye.*

## Limitations

Although the above tools and datasets provide a unique resource for predicting 2D heterostructures, there are various limitations that we mention here. First, the conventional semi-local DFT functionals we use in this work are known to underestimate the bandgaps of materials[58]. Higher levels methods such as hybrid-functionals[59, 60], $G_0W_0$ [61, 62] Bethe-Salpeter (BSE)[63] are needed to correct this issue, however, they are computationally very expensive. It is important to check the accuracy of the band-edge positions of the OptB88vdW with respect to HSE06 type methods. The OptB88vdW is taken from the JARVIS-DFT while the PBE and HSE06 data are taken from Ozcelik, et al.[15] for 14 materials such as BN, 2H-$MoS_2$, 1T-$HfS_2$. In Fig. 10, we compare the band-edges of 14 2D materials obtained from OptB88vdW with respect to PBE and HSE06 functionals. The Pearson correlation coefficient is used to find the relationship of data from different DFT methods. We find that the PBE and OptB88vdW band-edges adjusted with respect to the vacuum level of each material are very similar, with PC more than 0.96 for both CBMs and VBMs. This indicates that the conventional DFT method based heterostructure predictions should be comparable among themselves. Next, we compare the band-edges with respect to HSE06. We find that the CBMs of the two functionals are highly correlated while the VBMs are less correlated with deviations mainly attributed to nitride class of materials such as BN and AlN.

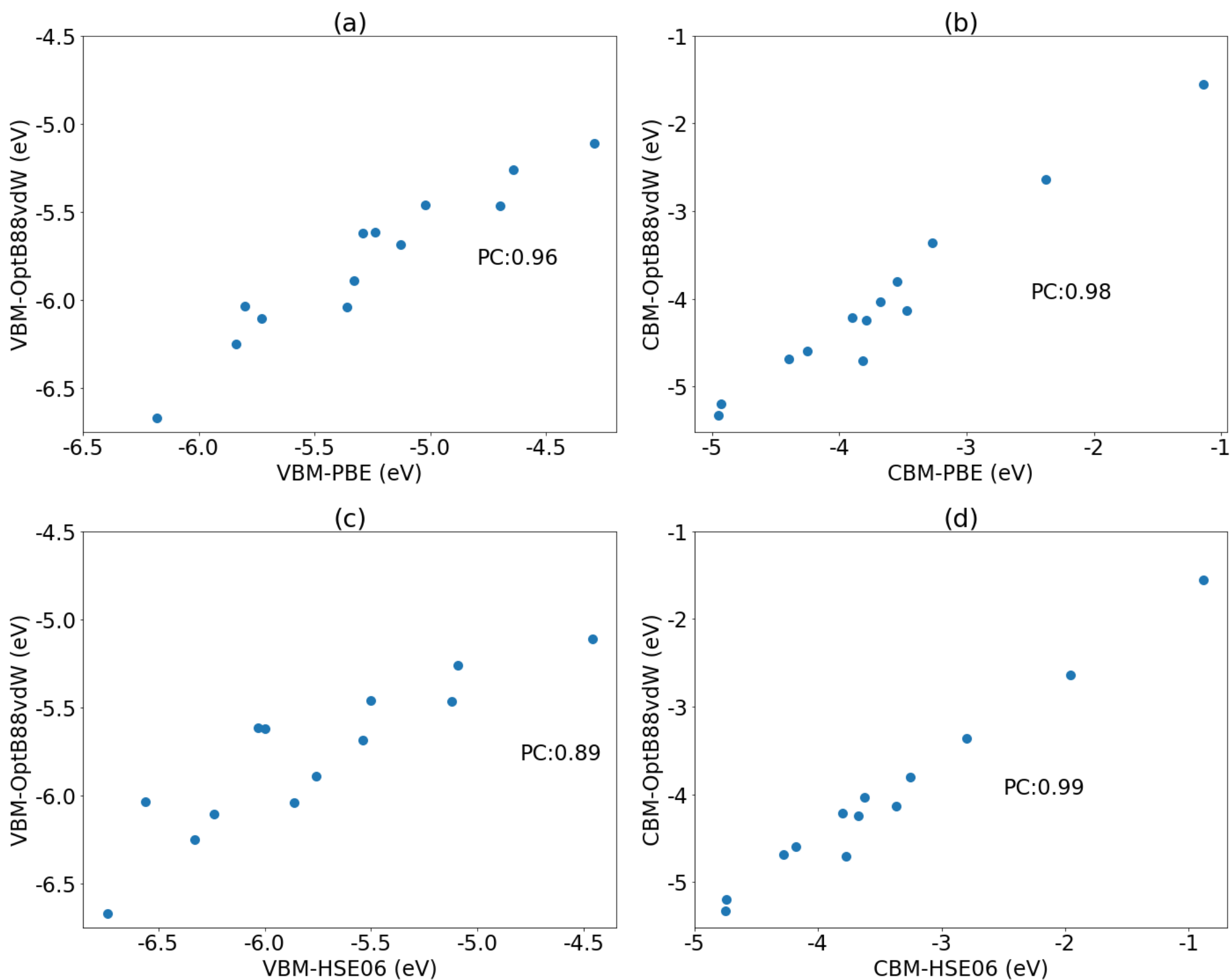

*Fig. 10 Parity plot comparing the VBM and CBM calculated with OptB88vdW, to those calculated with the PBE (a-b) and HSE06 (c-d) functionals. The Pearson Coefficient (PC) is a measure of correlation defined as covariance of two variables divided by both their standard deviations.*

In addition to the bandgap, it is important to consider the selective optical transitions in the vdW heterostructures, and carrier effective mismatches. For many of our 2D materials, we report the effective mass of individual 2D materials with the help of BoltzTrap code. Examples of using this code are given in our github page [https://github.com/JARVIS-Materials-Design/jarvis-tools-notebooks]. The effective masses of the both the 2D materials in a hetersotrucuture should be comparable because the carriers from the VBM/CBM of one material would migrate to the CBM/VBM of another and it is important that the carriers encounter a similar band-structure, hence also the similar effective masses[64]. Distribution of the effective mismatches of 2D heterostructures in Fig. 11 shows that the majority of the heterostructures have negligibly small effective mass-mismatches with the highest ones up to $4m_e$. This suggests that most of the heterostructures would have the effective mass compatibility. Note that computational data of effective masses for 2D materials are available for the majority of systems in our database, but for the systems where such information is not available, subsequent calculations could be necessary for heterostructure designs. It is also important to check if the optical transitions are allowed based on the symmetry of the 2D materials, and generally requires optical property calculations for heterostructures, which is beyond the scope of present work. For example, in Fig. 2c and Fig. 2d, the DFT calculations or band-diagrams can only provide the information on the existence of electronic states at a given energy. It does not provide the information on whether electronic transitions are allowed based on the symmetry of the wavefunctions. Nevertheless, we believe the tools and data would be useful for the initial selection of heterostructures.

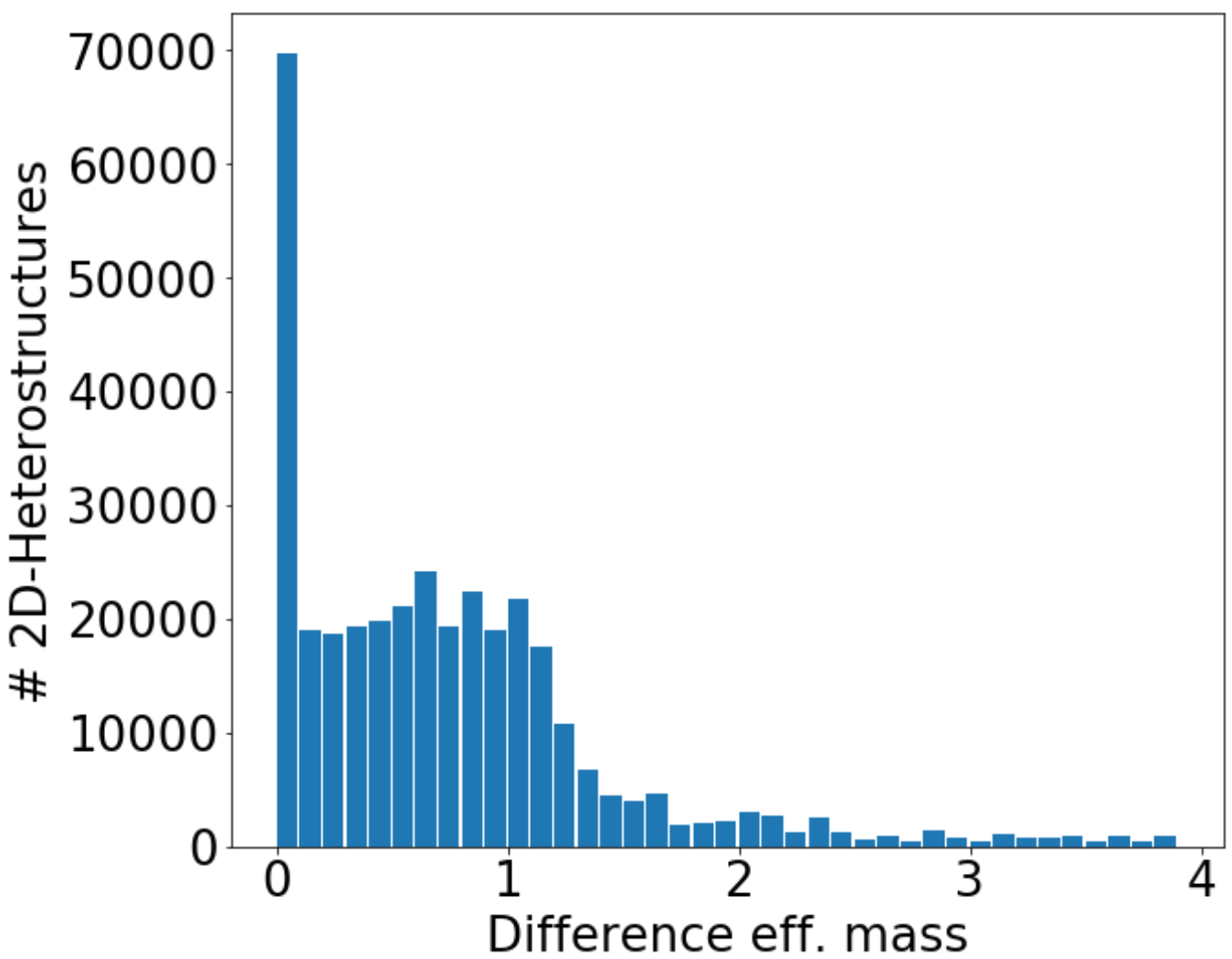

*Fig. 11 Histogram of the difference in effective mass between the two materials making up each 2D heterostructure in our study. The effective mass is in units of electron rest mass.*

In summary, we present a computational database, tools, web-apps and machine-learning models to accelerate the design and discovery of 2D-heterostructures. Using lattice-parameter and DFT based electronic level information of 674 non-metallic 2D-materials, we generate 226779 heterostructures and classify them into type-I, II and III systems according to Anderson's rule. We find that type-II is the most common and the type-III the least common heterostructure type. We analyze the chemical trends of all the generated heterostructures in terms of the periodic table of constituent elements and identify which atomic constituents are more likely to form low-mismatch and a particular type of heterostructures. The band alignment data can also be used for finding photocatalysts and high-WF 2D-metal contacts. We validate our results with respect to experimental data for a few systems and compare the electronic levels with respect to hybrid-functionals. Additionally, we carry out explicit DFT simulation of a few selected heterostructures, to compare the band-alignment description with the predictions from Anderson's rule. Finally, we develop web-apps and machine-learning models to enable users to virtually create combinations of 2D heterostructures and predict their properties.

## Methods

DFT calculations were carried out using the Vienna Ab-initio simulation package (VASP)[44, 45] software using the workflow given on our 'jarvis-tools' github page (https://github.com/usnistgov/jarvis ). Please note commercial software is identified to specify procedures. Such identification does not imply recommendation by National Institute of Standards and Technology (NIST). We use the OptB88vdW functional[46], which gives accurate lattice parameters for both vdW and non-vdW (3D-bulk) solids[35]. The crystal structure was optimized until the forces on the ions were less than 0.01 eV/Å and energy less than $10^{-6}$ eV. The CBM and VBM are determined based on DFT calculations of monolayers. Also, we calculate the local potential containing ionic plus Hartree and local exchange contributions to determine the VAC of a 2D monolayer material. The VAC is subtracted from the VBM and CBM to enable the comparison of band-diagrams of individual 2D materials in band-alignment diagrams. We use the method of Zur et al.[33, 34] to generate heterostructures of these 2D materials and we calculate the corresponding mismatch of the lattice constants and angles. The implementation is available on our github package, which is adapted from previous works by Mathew et al[32] and Dwarkanath et al[33]. Our implementation of Zur et al.'s method allows easy integration of the database with the available computational-tools at our github page. For two given materials, the band-alignments are determined based on the DFT computations performed using the OptB88vdW method. Based on the band-alignment data their applications and types of heterostructures are determined. We require lattice-mismatch less than/equal to 0.05 % and 1-degree angle tolerance to build the heterostructures. However, we provide all the data and tools to enable users to construct heterostructures with other tighter or looser tolerances. For metallic systems, we determine the workfunction as the difference between the Fermi-energy and the VAC. The web-app is developed using Flask-python[47] and jarvis-tools packages. The Flask-app allows integration of python based-programs and web using very light coding requirements. Finally, we train three neural network models for CBMs, VBMs and work-functions, according to the ALIGNN system previously described elsewhere.[43] To optimize hyperparameters we train on 80% of the data and optimize the performance on the 10% validation set. The ALIGNN is trained for 200 epochs and appears to be fully converged at that point, as shown in the convergence plots of Figure 12. We begin with a default choice of 0.01 for both learning rate and decay rate, with a batch size of 32, and test varying these parameters separately by a factor of two in either direction. The validation accuracy proved to be relatively insensitive to this variation, so we adopt the highest-performing settings discovered this way but do not optimize any further. For CBM, accuracy is maximized with the defaults for each parameter, while VBM is slightly more accurate with a batch size of 64, and work-function is most accurate with a learning rate of 0.02. We also perform a five-fold cross-validation study on the whole data to analyze the sensitivity of the selection of dataset-splits.

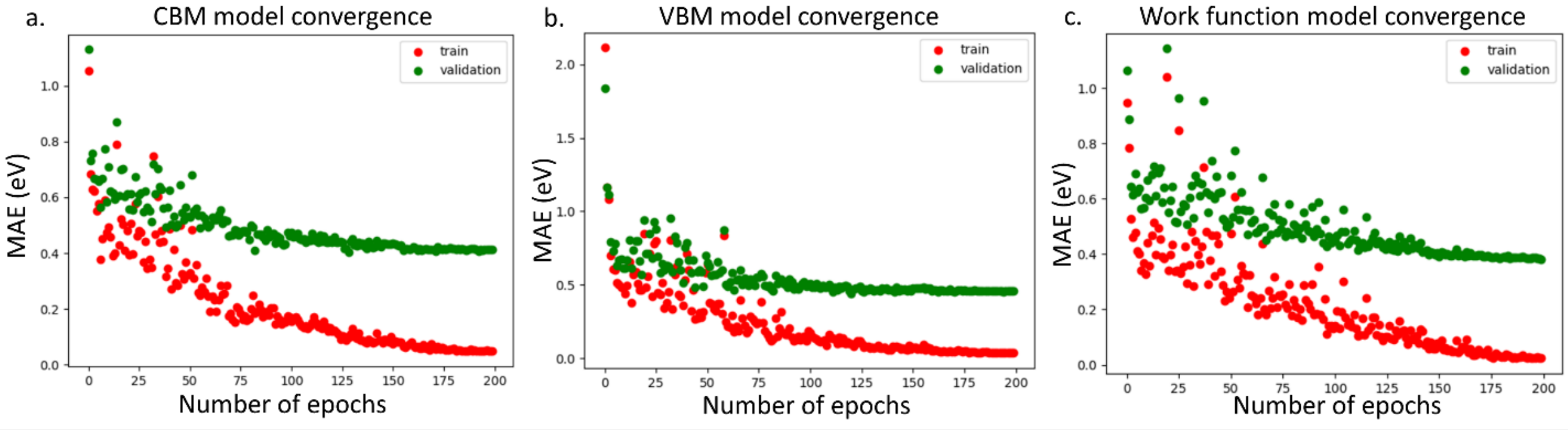

*Fig. 12 Convergence plot showing train set and validation set performance as a function of number of training epochs, for (a) ALIGNN CBM model, (b) VBM model, (c) WF model.*

## Data availability

The electronic structure data is available at the JARVIS-DFT website: https://jarvis.nist.gov/jarvish/ and [http://jarvis.nist.gov](http://jarvis.nist.gov) .

## Acknowledgments

K.C., K.F.G., and F.T. thank National Institute of Standards and Technology for funding, computational and data-management resources (NIST-Raritan, and NIST-CTCMS). K.C. also thank the computational support from XSEDE computational resources under allocation number (TG-DMR 190095). S. T. H. and G.P. would like to acknowledge support from the Los Alamos National Laboratory's Laboratory Directed Research and Development (LDRD) program's Directed Research (DR) project #20200104DR. Los Alamos National Laboratory is operated by Los Alamos National Security, LLC, for the National Nuclear Security Administration of the (U.S.) Department of Energy under contract DE-AC52-06NA25396.

## Author contributions

KC and KFG jointly developed the workflow. GP helped in the DFT-validation and machine learning based model development. STH trained and validated the machine learning model. KC carried out the DFT calculations and developed the web-app. All authors contributed in writing the manuscript and analyzing the results.

## Competing interests

The authors declare no competing interests.